\documentclass{aastex63}
\usepackage[utf8]{inputenc}
\usepackage{url}

\usepackage{color}

\usepackage [english]{babel}
\usepackage [autostyle, english = american]{csquotes}
\MakeOuterQuote{"}

\def\s{\phantom}

\def\amin{\ifmmode^{\prime}\else$^{\prime}$\fi}
\def\asec{\ifmmode^{\prime\prime}\else$^{\prime\prime}$\fi}

\def\simgt{\lower.5ex\hbox{$\; \buildrel > \over \sim \;$}}
\def\simlt{\lower.5ex\hbox{$\; \buildrel < \over \sim \;$}}

\newcommand\XTE{{\it RXTE\/}}

\newcommand\chandra{{\it Chandra}}

\newcommand\xmm{{\it XMM-Newton}}

\newcommand\swift{{\it Swift\/}}

\begin{document}

\title{Comment on "On the recurrence times of neutron star X-ray binary transients and the nature of the Galactic Center quiescent X-ray binaries"}

\author[0000-0002-9709-5389]{Kaya Mori} 
\affiliation{Columbia Astrophysics Laboratory, Columbia University, New York, NY 10027, USA}

\author[0000-0002-6126-7409]{Shifra~Mandel}
\affiliation{Columbia Astrophysics Laboratory, Columbia University, New York, NY 10027, USA}

\author{Charles~J.~Hailey}
\affiliation{Columbia Astrophysics Laboratory, Columbia University, New York, NY 10027, USA}

\author[0000-0002-7187-9628]{Theo~Y.E.~Schutt}
\affiliation{Columbia Astrophysics Laboratory, Columbia University, New York, NY 10027, USA}

\author[0000-0002-6395-4528]{Keri~Heuer}
\affiliation{Department of Physics, Cornell University 109 Clark Hall, Ithaca, New York 14853-2501, USA}

\author[0000-0002-1323-5314]{Jonathan~E.~Grindlay}  
\affiliation{Harvard-Smithsonian Center for Astrophysics, Cambridge, MA 02138, USA}

\author[0000-0002-6089-5390]{Jaesub~Hong} 
\affiliation{Harvard-Smithsonian Center for Astrophysics, Cambridge, MA 02138, USA}

\author[0000-0001-5506-9855]{John~A.~Tomsick}
\affiliation{Space Sciences Laboratory, University of California, Berkeley, CA 94720, USA}

\email{kaya@astro.columbia.edu}

\section{Introduction} 

In 2018, we reported our discovery of a dozen quiescent X-ray binaries in the central parsec (pc) of the Galaxy \citep{Hailey2018}.  In a recent follow-up paper \citep{Mori2021}, we published an extended analysis of these sources and other X-ray binaries (XRBs) in the central pc and beyond, showing that most if not all of the 12 non-thermal sources are likely black hole low-mass X-ray binary (BH-LMXB) candidates.  In response, \citet{Maccarone2022} (TM22 hereafter) 
argued, primarily on the claim  
that neutron star low-mass X-ray binaries (NS-LMXBs) often do not have short outburst recurrence times ($\simlt10$ yr), that they cannot be excluded as a designation for the 12 quiescent X-ray binary sources.  TM22 cites three main factors in their study: (1) X-ray outburst data of NS transients detected by \textit{RXTE} and \textit{MAXI}, (2) the Galactic population of NS-LMXBs, and (3) (persistently) quiescent NS-LMXBs in globular clusters.  
We address these arguments of TM22 and correct their misunderstandings of our work and the literature, 
even though most of these points have already been thoroughly addressed by \citet{Mori2021}. We also correct TM22's assertion that our arguments are based solely on NS transients' recurrence times. 
An executive summary is provided in \S2.  \S3-5 contain our responses to the three main arguments in TM22.  \S6 reiterates the conclusions, inferences and claims in \citet{Mori2021}.  Any comments are welcome and should be sent to the corresponding author (K.  Mori).   \\

\section{Executive summary}

(1) \textbf{The list of NS transients in TM22 is largely incomplete 
and highly biased, containing only some of the brightest X-ray outbursts (those exceeding all-sky monitors' detection limits); hence it does not represent a reliable sample of X-ray outburst recurrence times.}
 (see \S \ref{ns_transient_list}) \\

(2) \textbf{The recurrence time estimates presented in TM22 are inaccurate and highly model-dependent.} We found none of their arguments based on the (ambiguous or outdated) NS-LMXB or MSRP population studies compelling or even relevant to the non-thermal X-ray sources in the central pc (See \S \ref{ns_population}).   \\ 

(3) \textbf{Inferences on recurrence time drawn from globular clusters (the third point in TM22) are unreliable because:} (1) globular cluster environments (and stellar populations/evolutionary histories/dynamics) are very different from the GC, (2) X-ray observation coverage for globular clusters is far sparser compared to the GC region, which has been monitored nearly continuously for over two decades, and (3) the dearth of confirmed NS-LMXBs with quiescent \textit{non-thermal} X-ray emission (like that of the 12 central pc sources) in globular clusters (see \S\ref{globulars}).  \\

(4) \textbf{TM22 did not investigate the recurrence times of BH transients.}  We showed that the longer recurrence times observed in BH transients are more reliable than their NS counterparts, as BH X-ray outbursts are generally brighter -- and last longer -- than those of NS transients, and therefore are less likely to go undetected.  (See \S 6.2).  \\ 

(5) \textbf{Our BH-LMXB interpretation of the dozen non-thermal X-ray sources is supported by the X-ray transient data obtained in the GC; the GC sample contains the most complete and reliable measurements of recurrence times for different types of X-ray transients within a volume-limited region. } (See \S 6.1). \\

(6) \textbf{Contrary to the representation in TM22, we have never claimed that \textit{all} 12 non-thermal X-ray sources in the central pc must be BH-LMXBs. } Still, a BH-LMXB interpretation is most consistent with the data as well as theoretical predictions that predate our papers; it is also supported by an abundance of recent numerical simulations (See \S6.3). \\

\section{The analysis of NS transients in TM22 \S2 is incomplete and strongly biased} \label{ns_transient_list} 

The nature of X-ray outbursts is known to be stochastic.  X-ray fluxes, durations and recurrence times vary significantly between outbursts, often even from the same object.  
This stochasticity is an artifact of the ignition mechanism -- namely, accretion disk instabilities -- and results in X-ray outbursts that can last for as little as a few days, or as long as a year or more.  To avoid missing faint and/or short-duration X-ray outbursts, one  needs to monitor the region in question frequently and deeply.  For example, AX J1745$-$2901 is a known NS-LMXB located $\sim1.5$\amin\ from Sgr A*.  AX J1745$-$2901 undergoes repeated outbursts every few years, with peak X-ray fluxes that vary significantly between outbursts \citep{Degenaar2017}.  
TM22 makes no mention of such variations in flux or outburst duration, both of which strongly affect the detectability of X-ray outbursts and thus the \textit{observed} recurrence times.  Instead, they imply that any source that was at one point bright enough to be detected by 
an all-sky monitor must necessarily always have reached the same luminosity in prior or subsequent outbursts.  This is the given justification for ignoring the poorer pre-\textit{MAXI} sensitivity of all-sky monitors in establishing a longer baseline for the study of outburst recurrence, and for examples of "clustering" of outbursts.  
We were able to detect faint X-ray outbursts in the GC -- which would have been missed by all-sky X-ray monitors if they occurred elsewhere in the Galaxy -- only because \swift-XRT (often augmented by other X-ray telescopes) nearly continuously observes that region.  No other region in our Galaxy is monitored nearly as consistently.

In \citet{Mori2021} we referenced \citet{Carbone2019}, who showed that the GC provides a more complete sample of X-ray outburst data than any other region in our Galaxy.  We also cited \citet{Lin2019}, who addressed X-ray recurrence time differences between BH- and NS-LMXBs and clarified that NS-LMXBs are more vulnerable to the selection biases imposed by all-sky monitors' detection limits (i.e. they tend to have fainter and shorter X-ray outbursts, which are more easily missed).  However, as explained in \citet{Mori2021} (Section 3.1), the sheer quantity of exposure time and cadence frequency of the GC observations over the last two decades (especially the Swift-XRT monitoring over the last 15+ years) 
exceed every other region in our Galaxy by orders of magnitude.  Consequently, our detection sensitivity to much fainter X-ray outbursts within that region is greatly enhanced.  

TM22 acknowledges that the X-ray transient samples collected by all-sky X-ray monitors are limited to only bright sources:  "\textbf{Among the known samples (which are strongly biased toward sources bright enough to trigger all-sky instruments)}, typical black hole X-ray binaries show peak luminosities in their transient out- bursts of $10^{37-39}$ erg/s (Tetarenko et al. 2016a), while typical NS LMXBs show peak luminosities about a factor of 10 fainter (Yan \& Yu 2015)."  
Unfortunately, TM22 failed to explore the implications of this bias vis-\`a-vis \textit{detected} outburst recurrence times for NS transients.  It is reasonable to assume that the poorer detectability combined with the significantly higher \textit{observed} recurrence rate for NS transients (compared to their BH counterparts) implies that NS-LMXBs are indeed intrinsically more frequent outbursters.  Further confirmation of NS transients' \textit{intrinsic} outburst recurrence is provided by the aforementioned robust monitoring campaign of the GC, which is not subject to the same detection biases as the all-sky surveys.  All six identified NS transients in the GC have shown frequent outbursts (some of which would have been undetectable with all-sky monitors). 
\textbf{TM22 did not include what is arguably the most reliable recurrence time data of NS-LMXBs (including outbursts that exceeded their imposed flux limits) that was yielded by the \swift-XRT GC and Bulge monitoring programs (see \S3.1).  Instead, TM22 relied primarily on a catalog that is 15 years out of date, supplemented by a cherry-picked set of five more recent NS transients.}\footnote{TM22 states that they present a total of 21 long-recurrence sources, with six being added to a selection from the catalog by \citet{Liu2007}, but only 20 sources total are actually listed, including five additions to \citet{Liu2007}.}

\subsection{Recurrent NS-LMXBs not included in TM22}  

TM22 utilized a limited sample of NS-LMXBs for their analysis.  As they don't list the full set of 50 sources that they based their analysis on, it is uncertain which transients are accounted for in their work and which ones are not.

\subsubsection{NS transients in the GC}
 
Table \ref{tab:nsgc} lists the six NS transients in the GC, all of which have shown X-ray bursts and/or pulsations.  Though all six are frequent outbursters, the peak X-ray luminosities for individual transients often varied by several orders of magnitude, ranging between $\sim10^{34}-10^{37}$~erg/s.  Hence, some of these sources would have been classified as VFXTs at some point in their histories.  All six exhibited at least one outburst that reached or exceeded a peak $L_X$ of a few times $10^{36}$~erg/s.

\setlength{\tabcolsep}{15pt}
\begin{deluxetable*}{l c c c c}[h!]
\tablecaption{NS Transients in the GC}
\tablecolumns{5}
\tablehead{ \colhead{Name} & \colhead{Offset from   } &\colhead{Outbursts} & \colhead{$<t_{rec}>$\tablenotemark{b}}  & \colhead{Outburst dates\tablenotemark{c}} \\ [-5pt]
  & \colhead{Sgr A* (pc)} & \colhead{detected\tablenotemark{a}} & \colhead{(years)} }
\startdata   
AX J1745.6–2901 & 3.2 & 7 & $<3$ & '06, '07, '10, '13, '17, '19, '21    \\
GRS 1741.9–2853 & 23.3 & 9 & $<3$ & '05, '06, '07, '09, '10, '13, '16, '17, '20    \\
XMM J174457–2850.3 & 32 & 8 & $<3$ & '05, '08, '09, '10, '12, '13, '14, '16  \\
SAX J1747.0–2853 & 45.5 & 10 & $\simlt2$ & '01, '04, '05, '06, '07, '09, '11, '12, '16, '19  \\
GRO J1744–28 & 50.5 & 5 & $\simlt4$ & '08, '11, '14, '17, '21  \\
KS 1741–293 & 54.2 & 7 & $<3$ & '05, '08, '10, '11, '13, '16, '19  \\
\enddata
\tablecomments{The detection efficiency of outbursts varies spatially (being somewhat poorer at larger offsets from Sgr A*) and over time.  Though monitoring of the GC region has been frequent for over two decades, it was somewhat less consistent prior to 2006; hence, the number of outbursts (and mean recurrence time) should be taken as a lower (upper) limit.}
\tablenotetext{a}{Outbursts detected prior to 2001 are not included in this tally.}
\tablenotetext{b}{Mean recurrence time is calculated from the number of outbursts over the two-decade period between 2001-2021.}
\tablenotetext{c}{These dates were collated through a search of Astronomer's Telegrams; some may have been missed.}
\label{tab:nsgc}
\end{deluxetable*}

It is unknown whether any of the above NS transients were included in the analysis by TM22.  But there is no doubt that they comprise the most complete and unbiased record of X-ray outbursts.  
Since all-sky monitors do not detect fainter X-ray outbursts below the \textit{MAXI}/\XTE\ sensitivity limits, \textbf{it is impossible to distinguish between NS-LMXBs with intrinsically long recurrence time and those with a mixture of (infrequent) bright outbursts and fainter ones outside the GC. } In contrast, the NS-LMXB outbursts detected by more sensitive and frequent \swift-XRT observations of the GC -- the only region with reliable outburst data unbiased by luminosity constraints -- suggest that NS-LMXBs are predominantly frequently recurrent outbursters.  \\

\subsubsection{NS transients outside the GC}

(Note: we requested the list of 45 NS LMXBs from the corresponding author of TM22.  We will update this section if and when we receive it.) \\

TM22 based their analysis on a set of 45 NS LMXBs from \citet{Liu2007}, to which they added five more recent transients that have only once been observed in outburst.  Given that TM22 does not reveal the full set of 50 sources that their analysis is based on, it is impossible to fully evaluate its completeness, but the age of the catalog -- and the omission of any more recently discovered NS-LMXBs with more than one outburst --  is puzzling.  
Table \ref{tab:nsMW} lists a collection of recurring NS transients, 
which were not listed in TM22.  The inclusion of these sources in the analysis of TM22 is unknown, except for those transients listed in the lower part of the table (separated by the horizontal line), which are not listed in \citet{Liu2007} or in TM22 and therefore were not included.  

Many of these sources have not been detected in outburst as frequently as their GC counterparts, but it is nearly impossible to determine whether they have intrinsically longer recurrence times (possibly an artifact of lower accretion rates in older systems -- see \S3.2) or whether a significant number of outbursts went undetected due to the sensitivity limitations of all-sky monitors and limited targeted observations.  Although we consider the GC data far more reliable and unbiased, not to mention more applicable to our area of interest (given the varying stellar populations and ages of the GC vs the rest of the Galaxy), we still find it instructive to analyze those sources.

\setlength{\tabcolsep}{20pt}
\begin{deluxetable*}{l c c}[h!]
\tablecaption{NS Transients in the GC}
\tablecolumns{5}
\tablehead{ \colhead{Name}  &\colhead{Outbursts} & \colhead{Outburst dates\tablenotemark{b}} \\ [-5pt]
  & \colhead{detected\tablenotemark{a}} & \colhead{(years)} }
\startdata   
XTE J1739-285 & 4 & '05, '12, '19, '20 \\	
1A 1744-361 & 6 & '03, '04, '05, '08, '09, '13 \\
XTE J1751-305 & 4 & '02, '05, '07, '09 \\
IGR J00291+5934 & 3 & '04, '08, '15 \\
SAX J1750.8-2900 & 7 & '97, '01, '08, '11, '15, '18, '19 \\
4U 1724-307 & 3 &'96, '04, '08 \\
4U 1608-52 & $>15$ & 	\\	
MXB 1730-33 & $>30$ & $\sim$~bi-annual \\
Aql X-1 & $>11$ & \\
EXO 1745-248 & 3 & '00, '11, '15 \\
RX J1709.5-2639 & 9 & '97, '02, '04, '07, '10, '12, '13, '17, '20 \\
SAX J1808.4-3658 & 9 & '96, '98, '00, '02, '05, '08, '11, '15, '19 \\
GRS 1747-312 & $>30$ & $\sim$~bi-annual \\  
SAX J1748.9-2021 & 7 & '98, '01, '05, '09, '15, '17, '21 \\
XTE J1709-267 & & every $\sim2-3$~years \\
\hline
Swift J1756.9-2508 & 4 & '07, '09, '18, '19	\\
MAXI J0556-332 & 4 & '11, '12, '16, '20 \\
XTE J1810-189 & 3 & '08, '13, '20 \\
IGR J17511-3057 & 2 & '09, '15 \\
IGR J17407-2808 & 3 & '04, '11, '20 \\
IGR J17445-2747 & 2 & '17, '19 \\ 
CXOG1b J174852.7-202124 & 3 & '09, '10, '10\\ 
\enddata
\tablecomments{Sources listed above the horizontal line are found in \citet{Liu2007}, but may or may not have been included in the set of 45 transients used by TM22.  Sources listed below the line are not in \citet{Liu2007} and were not considered for TM22.}
\tablenotetext{a}{Outburst data was collated through a rough search of the literature; some reported outbursts may have been overlooked.  An unknown number of outbursts could have gone undetected.  Therefore, these numbers should be taken strictly as lower limits.}
\tablenotetext{b}{These dates were collated through a rough search of the literature; some may have been missed.}
\label{tab:nsMW}
\end{deluxetable*}

We note that three of the sources listed in Table \ref{tab:nsMW} (Aql X-1, 4U 1608-52, and MXB 1730-33) are highlighted by TM22 as being responsible for the majority of observed NS transient outbursts listed by \citet{Yan2015}.  However, \citet{Yan2015} also relied on a limited dataset -- namely, the outbursts detected by \XTE\ throughout a $\sim15$~year period.  This leaves out any outbursts that either occurred outside the \XTE\ monitoring area or were not luminous enough to be detected by \XTE; a significant fraction of outbursts that \textit{have} been detected (with other telescopes) would fall into the latter category.  For example, RX J1709.5-2639 is listed by \citet{Yan2015} as having had 3 outbursts, but we count at least 9, all but 2 preceding the publication of \citet{Yan2015}.  1A 1744-361 is listed as having 1 (\XTE-dectected) outburst, but we find six (including 4 detected by \XTE; ATel \#2305).  So while the above three sources are indeed responsible for a large fraction of the outbursts described in \citet{Yan2015}, there were many more outbursts from the other NS-LMXBs listed that were not included in the given tally.

TM22 claims that two of the three aforementioned sources (Aql X-1 and 4U 1608-52) "can both be inferred to have subgiant donor stars based on their orbital periods of 18.9 and 12.9 hours, respectively."  This is a rather bold inference, considering that stars with main sequence donors can have orbital periods of up to 48 hours \citep{Lin2019}.  But putting aside the question of whether these sources have evolved donors (and consequently, much higher mass accretion rates and more frequent outbursts), TM22 did not exclude sources (Cen X-4, XTE J1701-462, XTE J1723-376, and KS 1731-260) with periods believed or suspected to exceed 12.9 hours -- and in some cases, days -- from their list of long-recurrence NS-LMXBs.  
We also find that at least five of the 20 "long-recurrence" sources listed in Table 1 of TM22 (EXO 0748–676, KS 1731-260, 4U 1905+000, HETE J1900.1-2455, and GS 1826-238) are actually quasi-persistent (typically with $L_{\rm X} \sim 10^{35}-10^{36}$ erg\,s$^{-1}$)
, exhibiting "outbursts" or just $\sim$order-of-magnitude variations on timescales of decades, and hence are not truly "transient".  

\subsection{The cadence of X-ray outbursts}

TM22 (Section 2) states: "There is considerable and clear evidence that some X-ray binaries and cataclysmic variables show clustering in the times of their outbursts.", indicating that the observed short recurrence times could be due to clustering of X-ray outbursts. There is no reference cited for this "considerable and clear evidence" of clustering, but "Cen X-4, which showed extremely bright outbursts in 1969 and 1979," is given as "a good example."  Of course, given the dearth of continuous monitoring of X-ray transients outside the GC, we cannot rule out that such clustering indeed occurs, but the timescales of such clustering certainly cannot be determined given the varying outburst luminosities and detection sensitivities of all-sky monitors over different times and areas in the sky.  We can assert with moderate certainty that Cen X-4 likely did not undergo another "extremely" bright outburst since 1979, but we have no way of evaluating the potential number of not-extremely-bright outbursts that may have occurred in the intervening decades -- or the ones before. 

Despite the omission of BH transients from their analysis, TM22 bizarrely compares recurrence timescales of NS transients to that of classical and recurrent novae.  A nova outburst occurs when the hydrogen-rich envelope accreted onto the surface of a white dwarf (WD) in a cataclysmic variable (CV) binary system erupts in a thermonuclear runaway.  In contrast, LMXB outbursts are ignited by instabilities in the accretion disk surrounding a BH or NS.  Despite the differences between these sources and their outburst mechanisms, a careful analysis of the evolution of CVs is nevertheless illuminating
.  Long-term simulations of CV evolution show that as the donor mass is depleted, the mass transfer rate $\dot M$ drops significantly, with a corresponding decrease in the frequency of nova eruptions \citep{Hillman2020}.  Order-of-magnitude changes can occur on timescales of Myr or less.  It is plausible that a similar reduction in accretion rate and outburst frequency occurs over time for LMXBs -- hence, one might expect a population of semi-detached, persistently quiescent NS-LMXBs throughout the Galaxy, especially in regions with older stellar populations, such as the Galactic bulge and globular clusters (more in \S4-5).  However, as we remarked in \citet{Mori2021} (Section 3.1, paragraph 3; Section 6, \#6), \textbf{such low-accretion systems would have predominantly non-varying \textbf{thermal} spectra; they would appear nothing like the distinctly \textbf{non-thermal} (and mostly variable) sources we've identified as BH-LMXB candidates in the central pc.  It is unfortunate that this distinction was lost on TM22. } \\

\section{The Galactic population of NS-LMXBs discussed in TM22 \S3 is highly model-dependent and irrelevant to 
our work on the GC X-ray binaries} \label{ns_population}

Section 3 in TM22 argues that the recurrence time for NS-LMXBs can be inferred from population studies of 
NS-LMXBs and MSRPs. These points are irrelevant to our arguments as applied to our sources in the central pc, \textbf{as an underlying population of quiescent NS-LMXBs, however large, would not exhibit the distinctly non-thermal spectra observed from our sources (\S3.3).   
However, as we describe parenthetically below, 
this attempt for estimating the recurrence time for NS-LMXBs (or any class of object) is also entirely dependent on 
rather crude and poorly constrained models. }

\begin{itemize} 

\item TM22 explains that according to population synthesis models, $2,000-10,000$ NS-LMXBs should exist in the Milky Way.  This is an overall estimate and it is likely that the different environments in various regions of the Galaxy, such as the Galactic Center and globular clusters, spur their own unique binary formation and evolution paths.  This implies potentially diverging NS-LMXB population densities between different regions, not to mention disparate age distributions of binaries (with strong implications for $\dot M$ and outburst recurrence -- see \S3.3). 
Apart from the factor of $\sim5$ uncertainty in the overall number of NS-LMXBs, such an estimate for the entire galaxy 
provides only vague implications for any specific region, including the GC, 
which is distinct (in environment and stellar population) from the Galactic disk and from globular clusters.  Given the lack of sensitivity to faint outbursts outside the GC, this hardly signifies that "the total number of NS X-ray binaries [is] underpredicted by a factor of $\sim50$" -- even if we assume that the underlying population of NS-LMXBs only includes transient systems.  The same is true of the characteristic lifetime estimate for LMXBs based on the projected number of MSRPs (see below). 

\item TM22 states, "The total MSRP population is about 40,000 (Lorimer 2008)."  However, \citet{Lorimer2008} stated (emphasis ours), "Integrating the local surface densities of pulsars over the whole Galaxy requires a knowledge of the presently rather \textbf{uncertain} Galacto-centric radial distribution.  One approach is to \textbf{assume} that pulsars have a radial distribution similar to that of other stellar populations and to scale the local number density with this distribution in order to estimate the total Galactic population.  The corresponding local-to-Galactic scaling is $1000\pm250$ kpc$^2$.  This implies a population of $\sim$160,000 active normal pulsars and $\sim$40,000 millisecond pulsars in the Galaxy."  According to this quote, not only the number but also the distribution of MSRPs in our Galaxy is 
highly uncertain and subject to arbitrary assumptions, 
as indicated by 
the lack of MSRP detections in the GC.  Once again, the limited detection sensitivity outside the GC renders the point irrelevant.  

\item TM22 states (emphasis ours), "there should be 100-500 NS LMXBs within 3 kpc \textbf{if the population synthesis estimates are correct.}"  They fail to mention that this prediction also relies on the assumption that NS-LMXBs follow the overall stellar mass distribution in the Galaxy.  They do not elaborate on how many of these predicted NS-LMXBs are X-ray emitting binaries with (1) Roche lobe-filling donors, (2) a mass accretion rate in the range that can produce X-ray outbursts, (3) non-thermal quiescent X-ray emission (such as produced by our central pc sources), and (4) detectable pulsations and/or type I X-ray bursts (to allow for NS identification).  
As far as we know, the only plausibly robust 
study of the number densities of NS-LMXBs and MSRPs 
is the recent N-body simulation of the globular cluster 47 Tuc \citep{Ye2021}.  We are not aware of any concrete predictions with similar levels of plausibility for the entire Galaxy, the GC, or the Bulge.  Any recurrence time estimates based on the population studies are highly model-dependent and subject to large uncertainties.  As TM22 states (emphasis ours), "The range of estimates from first principles binary population synthesis depends largely on treatment of natal kicks of the NS and on the common envelope phase of binary evolution, 
\textbf{both of which are relatively poorly constrained from the existing data, and are extremely challenging problems to approach using theoretical methods.}"  
(Again, we have no reason to dismiss the possibility that there is a significant population of undiscovered NS-LMXBs that undergo no outbursts on timescales of decades or longer because they are accreting at very low rates; but such systems would not have the non-thermal emission observed from our central pc sources.) 

\item Additionally, TM22 used only $\sim5$ NS-LMXBs with known distance estimates of $<3$~kpc to estimate limits on the recurrence time of local NS-LMXBs.  Note the 3rd footnote in TM22 stating, "The number is approximate because there are many sources with distance uncertainties that allow them to be closer to or further than 3 kpc."  Given the small (and potentially flux-limited -- as even those distances may be poorly constrained!) sample of local NS-LMXBs, TM22's recurrence time estimates are likely inaccurate.  \textbf{Of course, this is also of limited relevance to our classification of the \textit{non-thermal} X-ray sources in the central pc, as explained above.} 

\item TM22 states that "binary MSRPs predominantly have orbital periods of at least 1 day (Lorimer 2008), and often have much longer orbital
periods."  This assessment is long outdated and is now known to be incorrect.  A substantial fraction of MSRPs, particularly those found in globular clusters, and especially "spider" systems, have periods of $\simlt1$~day (see e.g. Alessandro Patruno's MSP catalogue at \url{https://apatruno.wordpress.com/about/millisecond-pulsar-catalogue/}).

\end{itemize}

TM22's claims are grossly inconsistent with our current understanding of GC population studies. A plethora of theoretical and analytical models predict a large concentration of BH-LMXBs in the central pc \citep{Generozov2018, Panamarev2019, Tagawa2020, Gruzinov2020} or even preclude a significant number of NS-LMXBs in that region \citep{Bortolas2017}.  It is worth noting that the above models are arguably more robust and applicable, being particular to the central pc region -- unlike the more outdated population studies cited by TM22. \\

\section{The arguments based on NS-LMXBs in globular clusters (TM22 \S4) are poorly supported -- and irrelevant to the GC X-ray binary population} \label{globulars} 

There are a number of reasons why NS-LMXBs found in globular clusters make a very poor analog to those in the GC, as outlined below:

\begin{enumerate}
\item 
Globular clusters cannot provide an "apples-to-apples" comparison to the non-thermal X-ray sources in the central pc, because \textbf{hardly any confirmed 
NS-LMXBs in globular clusters show robust (quiescent) non-thermal 
emission.}  CX3 (EXO 1745-248) in Terzan 5 (at 8.7 kpc) is one such case which displays a hard, non-thermal spectrum in quiescence. However, the source has been detected in outburst in '00, '11, and '15.  Given that Terzan 5 is not frequently monitored with pointing observations, additional outbursts may well have been missed.  
Another source (Aql X-1), considered the "prototype" NS-LMXB, boasts a clear, variable, non-thermal spectral component and a recurrence time $\tau_{\rm rec} \simlt 1$~year. 
While non-thermal emission from a NS transient between outbursts is not entirely surprising, the same would not be expected in a persistently quiescent NS-LMXB. 
Most of the NS-LMXBs in globular clusters  
have thermal spectra, usually fit by blackbody or hydrogen atmosphere models, due to their low accretion rates (a likely consequence of their aged stellar populations).  
In contrast, we found that only one confirmed NS-LMXB -- CX2 (Swift J174805.3-244637) in Terzan 5 -- has a statistically robust non-thermal X-ray component in quiescence and only one detected X-ray outburst.  (Note that 
any number of fainter outbursts from this source could have easily gone undetected, as explained below.  Therefore, even CX2 is not evidence for long recurrence time in a non-thermal quiescent source.)

The \chandra\ coverage of X-ray binaries in globular clusters is very scarce compared to the GC with its $\sim7$~Msec total exposure.  As a result, even where included, the presence of a non-thermal (power law) component for NS-LMXBs (or candidates) is often not rigorously established due to poor statistics.  
To address the footnote in TM22 claiming that 
"Many of these thermal emitters [NS-LMXBs]
also show power law tails to their spectra," we examine some of those sources more closely below.
Based on a review of the literature, we find that both the
detection of a non-thermal component for (persistently) quiescent LMXBs in globular clusters, and the identification of such sources as NS-LMXBs, are highly questionable.  Following is a list of globular cluster NS-LMXB candidates we reviewed, complete with direct quotations from the literature (emphasis ours):

\begin{itemize}
    \item CXOU J180801.98$-$434255.3 in NGC 6541 (7.5 kpc): The MAVERIC survey paper \citep{Bahramian2020} states, "The power-law component is only loosely constrained, which sometimes suggests it may not be needed" and "The best-fit X-ray spectral model suggests that this system is \textbf{likely} an NS-LMXB accreting at low levels."  The power-law component does not appear to be significant. \\%
    
    \item CXOU J162740.51-385059.1 in NGC 6139 (10.1 kpc): The MAVERIC survey paper \citep{Bahramian2020} states that "These models indicate that the system is \textbf{probably} another weakly accreting NS-LMXB."  
    
    It is unclear from the information provided in the paper whether this power-law component is significant or well constrained, or whether the spectral component can be equally well modelled with a blackbody. \\

    \item CX3 in NGC 6440 (8.5 kpc): \citet{Heinke2003} states, "Our results strongly support the suggestion by Pooley et al. (2002) that CX2, CX3, CX5 and CX7 are likely qLMXBs.  \textbf{Although the best fits for CX3 and CX5 include a power-law component, this component cannot be considered significantly detected}."  It is also worth noting that CX2, CX3, CX5 and CX7 are all \textit{candidate} NS-LMXBs; their NS classification has not been confirmed \citep{Walsh2015}.  \\  
 
    \item CX1 (SAX J1748.9-2021) in NGC 6440: This source is one of the few confirmed NS-LMXBs in globular clusters.  It was detected in outburst in 1998, 2001, 2005, 2009-2010, 2015, 2017, and 2021 \citep{Sharma2019, Pike2021}.  It does exhibit a power-law component in quiescence, which is not surprising for an accreting and frequently outbursting source.  Similarly robust non-thermal quiescent emission is generally not observed from persistently quiescent  NS-LMXBs.  
\end{itemize}

In summary, the footnote in TM22 referencing "power law tails" for (persistently) quiescent NS-LMXBs is not supported by the evidence.  We 
believe that a more reliable comparison to the 12 non-thermal sources in the central pc can be made using
the behavior of \textit{confirmed} rather than \textit{candidate} NS-LMXBs (particularly those in the GC, who share a similar environment in a nearby region and whose outburst recurrence is well determined). \\

\item \textbf{X-ray outbursts with $L_X \simlt 10^{36}$ erg\,s$^{-1}$ from globular clusters are largely undetectable by all-sky monitors.}  
X-ray observations of \textit{any} globular cluster are far less extensive and sensitive compared to the ongoing X-ray monitoring of the GC.  Terzan 5, likely the most extensively observed globular cluster, boasts a total of $\simlt750$~ksec exposure \citep{Bogdanov2021}, compared to the GC with more than 7 Msec exposure from \chandra alone.  According to the HEASARC website (\url{https://heasarc.gsfc.nasa.gov}), 47 Tuc, 
another frequently observed target, 
has been observed 79 times (often within short intervals) by Swift-XRT, with exposures ranging from 45 to 6952 seconds, for a grand total of less than 195 ksec (or several orders of magnitude less than the GC \swift-XRT, \xmm, and \chandra\ surveys).   Other globular clusters have even poorer monitoring records.  

Since the distances to many globular clusters are comparable to the GC distance (8 kpc), the typical detection threshold for X-ray outbursts by all-sky monitors ($L_X \sim 10^{36}$ erg\,s$^{-1}$) holds for both regions.  In addition, X-ray transients detected in crowded regions (like globular clusters or the GC) by all-sky X-ray monitors or even Swift-XRT need to be localized by follow-up \chandra\ observations for firm source identification.  
\textbf{Given the limited sensitivity of all-sky X-ray monitors, fainter X-ray outbursts with $L_X\simlt10^{36}$~erg\,s$^{-1}$ have likely been missed.}  (In fact, several candidate NS-LMXB transients in globular clusters were not detected by all-sky monitors and their outbursts are only "on the record" because they serendipitously occurred during a pointing observation).  
Such objects could 
avoid detection by Swift/BAT or \textit{MAXI} in other regions of our galaxy, and there are 
multiple possible examples in globular clusters (although it is impossible to determine whether they represent a large population or not, given the incomplete X-ray monitoring outside of the GC region where the Swift-XRT detection sensitivity reaches $L_{\rm X} \sim 10^{34}$ erg\,s$^{-1}$).  However, as mentioned above, the X-ray observations of globular clusters are far less extensive and sensitive compared to the ongoing X-ray monitoring of the GC with Swift-XRT.  
And the GC X-ray transient population shows unambiguously that non-thermal (i.e. actively accreting) NS-LMXBs undergo frequent outbursts, while their BH counterparts do not; and had any of the 12 non-thermal sources in the central pc undergone even a faint outburst, it likely would have been detected, given the superior sensitivity limits in the GC region.

\item \textbf{The environments of globular clusters and the GC are different:}  
The GC and globular clusters comprise entirely different stellar populations and environments \citep{Heinke2003}, with different ages, metallicities, star formation histories, stellar densities and gravitational potential wells (i.e. the  presence vs lack of a supermassive BH in the center).
Unique X-ray binary formation mechanisms have been proposed for the GC \citep{Generozov2018}.  
For this reason, as well as the uniquely consistent X-ray observations of the Galactic Center, we consider it far more appropriate to compare our 12 central pc sources to the known LMXBs in the GC, ensuring an equivalent comparison. 

\end{enumerate}

TM22 counters our empirical arguments based on GC transients with globular cluster sources, but \textbf{there are hardly any examples that can support their position}.   
\textbf{For the reasons described above,
we believe that only the GC region will offer a complete sample for a valid comparison of X-ray transients}.  (Note that as a result of the far more extensive X-ray monitoring, the 12 sources in the central pc boast far better photon statistics for their quiescent X-ray spectra, as shown in \citet{Mori2021}).   \\

\section{Reiterating the major points in Mori et al.  2021 that were omitted or misunderstood by TM22} 

In their introduction, TM22 makes the incorrect claim that our argument for the BH-LMXB designation for our central pc sources is based solely on the assumption that NS transients have short recurrence times, and that we dismiss any long-recurrence systems as "irrelevant outliers."   In fact, \citet{Mori2021} analyzed a combination of factors that in sum, led us to conclude that the most plausible classification is BH- rather than NS-LMXBs.  Chiefly among those are:

\begin{itemize}
    \item The near-continuous monitoring of the GC region, 
    \item The distinct outburst histories of identified BH vs NS transients within this uniquely well-monitored region,
    \item The lack of 
    outbursts from those 12 sources in the central pc, and
    \item The distinctly non-thermal spectra. 
\end{itemize}

We also showed that the luminosity and spatial distributions of these sources is more consistent with an underlying population of BH-LMXBs [\citet{Mori2021}, \S3.3.1, \S3.2].  We augmented our analysis of the GC transients by examining the broader distributions of Galactic BH- and NS-LMXBs and how their overall outburst durations, frequencies, and peak luminosities diverge, in the context of the limited sensitivity of all-sky monitors [\citet{Mori2021}, \S3.1].  Our empirical analysis is additionally supported by a plethora of analytical models and simulations that predict a high density of BH binaries in the central pc [\citet{Mori2021}, \S5].  In the following paragraphs, we elaborate on the points listed above: \\

\subsection{The GC: unique among the regions in our Galaxy}

\paragraph{Reliability of outburst recurrence data based on detection efficiency}
In \citet{Mori2021}, we explained at length that our conclusions are primarily based on the observed NS and BH transients in the GC and their X-ray spectral/timing properties.  This is the most robust comparison of different X-ray source types as it concerns the same region where we detected the non-thermal X-ray sources.  Just as importantly, the GC has been much more frequently monitored by X-ray telescopes than any other region in our Galaxy.   
Given their limited sensitivity, transient data obtained only with all-sky X-ray monitors is not reliable for purposes of discerning outburst recurrence times \citep{Carbone2019}.  This is particularly true for NS transients, which tend to have fainter and shorter outbursts compared to their BH counterparts and are therefore more likely to go undetected when they outburst \citep{Yan2015, Lin2019}.  TM22 concedes this point, writing that "Among the known samples (which are strongly biased toward sources bright enough to trigger all-sky instruments), typical black hole X-ray binaries show peak luminosities in their transient out- bursts of $10^{37-39}$ erg/s (Tetarenko et al. 2016a), while typical NS LMXBs show peak luminosities about a factor of 10 fainter (Yan \& Yu 2015)."  But the GC region is not subject to those detection biases, because it has been extensively monitored with direct observations -- to much higher depth than that offered by all-sky monitors.

But limited flux sensitivity is not the only factor hobbling all-sky monitors in detecting NS transients; 
in globular clusters, where source crowding is significant, follow-up \chandra\ observations of X-ray transients are often necessary in order to identify outbursting sources discovered by all-sky monitors, given the latter's poor spatial resolution.  
Such follow-up observations are not always performed even when an outburst is detected by an all-sky monitor.  
By contrast, even faint and short-duration transient outbursts in the GC are typically detected and their source identified, thanks to the superior monitoring in the GC region.
As we explained in \citet{Mori2021}, this is why we
based our analysis on 
the GC X-ray transient data; it is simply the most complete, unbiased sample of outburst recurrence available \citep{Carbone2019}.  
No other region could provide a valid basis for comparison to the 12 non-thermal sources in the central pc.

\paragraph{The X-ray transient data in the Galactic Center}
Six NS-LMXBs (identified through type I X-ray bursts and/or pulsation detection) and four BH transients (identified through their spectral/broadband outburst properties) 
comprise the sample of identified X-ray transients in the GC.
Among this sample of LMXBs -- the most complete, relevant, and unbiased one available to us -- the distinction between NS and BH transient recurrence could not be sharper.  \textbf{Every single one of the NS transients showed short recurrence times (ranging on average between $2-4$~years [Table \ref{tab:nsgc}]); every BH candidate showed only a single outburst.}  It is therefore obvious that the 12 non-thermal sources in the central pc are consistent with BH-LMXBs, and not the NS-LMXB population, within the same region.  Indeed, it would be astonishing if all or most of the 12 sources 
turned out to be NS-LMXBs with outburst behaviors that are entirely inconsistent with that of every other identified NS-LMXB within $\sim55$~pc.

\paragraph{Spatial distributions of identified X-ray transients in the GC}
The four BH transients in the GC are all located 
in/near the central pc region, just like the 12 non-thermal sources.  
On the other hand, the six NS-LMXBs in the GC are widely distributed out to $\sim55$~pc.  The distinct spatial distributions of the identified BH and NS transients in the GC support the BH-LMXB interpretation in \citet{Hailey2018, Mori2021}, independently of ourtburst recurrence data.  

\paragraph{Unique environmental impacts on binary formation and evolution}
The central pc region is a different environment from the solar neighborhood, and distinct from globular clusters, since its star/binary formation is influenced by the supermassive BH's gravity.  The broader stellar population of the GC is substantially younger than that of most globular clusters, a factor that holds significant implications for accretion rates, and therefore, outburst recurrence timescales.  Additionally, unique binary formation mechanisms have been suggested for the GC region \citep{Generozov2018}.  Given those differences, any comparisons drawn between globular clusters and the GC would be of dubious validity. \\

\subsection{Typical recurrence timescales, outburst luminosity and duration for Galactic BH vs NS transients }

One of the major arguments in \citet{Mori2021} is that BH transient recurrence time data are more reliable than those of NS transients \citep{Lin2019}.  This is because, as TM22 states (emphasis ours), "Among the known samples \textbf{(which are strongly biased toward
sources bright enough to trigger all-sky instruments)}, typical black
hole X-ray binaries show peak luminosities in their transient outbursts
of $10^{37-39}$ erg/s (Tetarenko et al.  2016a), while \textbf{typical NS LMXBs show peak luminosities about a factor of 10 fainter} (Yan \& Yu 2015)."   
As we also point out in \citet{Mori2021}, NS transient outbursts also tend to be shorter, on average, than BH outbursts, further reducing their likelihood of detection compared to BH systems.
Despite the poorer average detectability of NS vs BH transient outbursts, several studies suggest longer recurrence times for BH transients \citep{Yan2015, Corral2016}.  
But TM22 did not explore the evidence for 
longer BH recurrence times.  
The former have longer average recurrence times than the latter, despite the highly biased TM22 selection criteria, which strongly favor BH outbursts.   

We did not claim -- as TM22 suggests -- that \textit{all} NS-LMXBs must necessarily undergo frequent outbursts.  To the contrary, we explicitly presented the possibility that a population of undiscovered persistently quiescent NS binaries exists, perhaps even in the GC.  We pointed out, however, that possible sources of that nature, detected in globular clusters, show predominantly thermal emission, as one might expect from a binary undergoing little to no accretion. \\

\subsection{Other errors and omissions in the description of our work given by TM22}

\paragraph{We did \textbf{not} rule out the 
possibility that a fraction of the 12 sources are not BH-LMXBs}

We stated that one/three of the dozen soft non-thermal sources could potentially be an SS-Cyg-like CV/MSRPs, respectively.  Yet, we consider the BH-LMXB population most plausible 
for most of these non-thermal X-ray sources after taking into account our spectral/timing results, the most updated 
data on X-ray transients, and simulations of the stellar/binary populations in the GC \citep{Mori2021}.  Like many scientific papers, we offered a plausible solution and made a reasonable hypothesis based on the current observations/knowledge, given the \chandra\ data sets, \swift-XRT monitoring and other X-ray transient data.  In \citet{Mori2021}, we made 
a robust case for why these results are most consistent with BH-LMXBs.  \textbf{While we do not completely rule out an alternative interpretations for a subset of our sources, these alternatives are simply less probable than the BH interpretation.  And while there is little to no theoretical/empirical support for those alternative suggestions, there is a wealth of evidence to support the BH designation.}  

\paragraph{Our BH-LMXB interpretation is consistent with a number of theoretical models and numerical simulations} 

The record is rife with theories and simulations that predict a high concentration of stellar-mass BHs (and BH-LMXBs) in the central pc of the Galaxy \citep{Morris1993, Generozov2018, Panamarev2019, Gruzinov2020, Baumgardt2018, Tagawa2020}.  The evidence for a similar concentration of NS-LMXBs in the same region is more murky, with some models suggesting a lower incidence of binary formation \citep{Generozov2018} or survival \citep{Bortolas2017} for NSs.  

\paragraph{Four of the 16 BH-LMXB candidates in/near the central pc previously outbursted}
TM22 states that of the 16 objects we identify as BH candidates, "none has shown an X-ray outburst."  This is incorrect -- our analysis involved 12 non-thermal sources that have never been detected in outburst, and four previously identified BH transients in/near the central pc.  The latter did undergo one bright outburst each and were identified as BH candidates based on either spectral properties or bright radio jets.

\paragraph{Definition of "bright" X-ray outbursts}
TM22 claims in their introduction that we failed to "define a large outburst quantitatively."  As a matter of fact, we didn't refer to any outbursts as "large," but we did quantify faint ($<10^{36}$~erg\,s$^{-1}$; in the Introduction, \S3.2, \S3.4) and bright ($>10^{36}$~erg\,s$^{-1}$; \S3.3) outbursts \citep{Mori2021}.


\begin{thebibliography}{}
\expandafter\ifx\csname natexlab\endcsname\relax\def\natexlab#1{#1}\fi

\bibitem[{{Bahramian} {et~al.}(2020){Bahramian}, {Strader}, {Miller-Jones},
  {Chomiuk}, {Heinke}, {Maccarone}, {Pooley}, {Shishkovsky}, {Tudor}, {Zhao},
  {Li}, {Sivakoff}, {Tremou}, \& {Buchner}}]{Bahramian2020}
{Bahramian}, A., {Strader}, J., {Miller-Jones}, J. C.~A., {et~al.} 2020, \apj,
  901, 57

\bibitem[{{Baumgardt} {et~al.}(2018){Baumgardt}, {Amaro-Seoane}, \&
  {Sch{\"o}del}}]{Baumgardt2018}
{Baumgardt}, H., {Amaro-Seoane}, P., \& {Sch{\"o}del}, R. 2018, \aap, 609, A28

\bibitem[{{Bogdanov} {et~al.}(2021){Bogdanov}, {Bahramian}, {Heinke}, {Freire},
  {Hessels}, {Ransom}, \& {Stairs}}]{Bogdanov2021}
{Bogdanov}, S., {Bahramian}, A., {Heinke}, C.~O., {et~al.} 2021, \apj, 912, 124

\bibitem[{{Bortolas} {et~al.}(2017){Bortolas}, {Mapelli}, \&
  {Spera}}]{Bortolas2017}
{Bortolas}, E., {Mapelli}, M., \& {Spera}, M. 2017, \mnras, 469, 1510

\bibitem[{{Carbone} \& {Wijnands}(2019)}]{Carbone2019}
{Carbone}, D., \& {Wijnands}, R. 2019, \mnras, 488, 2767

\bibitem[{{Corral-Santana} {et~al.}(2016){Corral-Santana}, {Casares},
  {Mu{\~n}oz-Darias}, {Bauer}, {Mart{\'{\i}}nez-Pais}, \&
  {Russell}}]{Corral2016}
{Corral-Santana}, J.~M., {Casares}, J., {Mu{\~n}oz-Darias}, T., {et~al.} 2016,
  \aap, 587, A61

\bibitem[{{Degenaar} {et~al.}(2017){Degenaar}, {Wijnands}, {Reynolds},
  {Miller}, \& {Kennea}}]{Degenaar2017}
{Degenaar}, N., {Wijnands}, R., {Reynolds}, M.~T., {Miller}, J.~M., \&
  {Kennea}, J.~A. 2017, The Astronomer's Telegram, 10900, 1

\bibitem[{{Generozov} {et~al.}(2018){Generozov}, {Stone}, {Metzger}, \&
  {Ostriker}}]{Generozov2018}
{Generozov}, A., {Stone}, N.~C., {Metzger}, B.~D., \& {Ostriker}, J.~P. 2018,
  \mnras, 478, 4030

\bibitem[{{Gruzinov} {et~al.}(2020){Gruzinov}, {Levin}, \&
  {Zhu}}]{Gruzinov2020}
{Gruzinov}, A., {Levin}, Y., \& {Zhu}, J. 2020, \apj, 905, 11

\bibitem[{{Hailey} {et~al.}(2018){Hailey}, {Mori}, {Bauer}, {Berkowitz},
  {Hong}, \& {Hord}}]{Hailey2018}
{Hailey}, C.~J., {Mori}, K., {Bauer}, F.~E., {et~al.} 2018, \nat, 556, 70

\bibitem[{{Heinke} {et~al.}(2003){Heinke}, {Grindlay}, {Lloyd}, \&
  {Edmonds}}]{Heinke2003}
{Heinke}, C.~O., {Grindlay}, J.~E., {Lloyd}, D.~A., \& {Edmonds}, P.~D. 2003,
  \apj, 588, 452

\bibitem[{{Hillman} {et~al.}(2020){Hillman}, {Shara}, {Prialnik}, \&
  {Kovetz}}]{Hillman2020}
{Hillman}, Y., {Shara}, M.~M., {Prialnik}, D., \& {Kovetz}, A. 2020, Nature
  Astronomy, 4, 886

\bibitem[{{Lin} {et~al.}(2019){Lin}, {Yan}, {Han}, \& {Yu}}]{Lin2019}
{Lin}, J., {Yan}, Z., {Han}, Z., \& {Yu}, W. 2019, \apj, 870, 126

\bibitem[{{Liu} {et~al.}(2007){Liu}, {van Paradijs}, \& {van den
  Heuvel}}]{Liu2007}
{Liu}, Q.~Z., {van Paradijs}, J., \& {van den Heuvel}, E.~P.~J. 2007, \aap,
  469, 807

\bibitem[{{Lorimer}(2008)}]{Lorimer2008}
{Lorimer}, D.~R. 2008, Living Reviews in Relativity, 11, 8

\bibitem[{{Maccarone} {et~al.}(2022){Maccarone}, {Degenaar}, {Tetarenko},
  {Heinke}, {Wijnands}, \& {Sivakoff}}]{Maccarone2022}
{Maccarone}, T.~J., {Degenaar}, N., {Tetarenko}, B.~E., {et~al.} 2022, \mnras,
  512, 2365

\bibitem[{{Mori} {et~al.}(2021){Mori}, {Hailey}, {Schutt}, {Mandel}, {Heuer},
  {Grindlay}, {Hong}, {Ponti}, \& {Tomsick}}]{Mori2021}
{Mori}, K., {Hailey}, C.~J., {Schutt}, T. Y.~E., {et~al.} 2021, \apj, 921, 148

\bibitem[{{Morris}(1993)}]{Morris1993}
{Morris}, M. 1993, \apj, 408, 496

\bibitem[{{Panamarev} {et~al.}(2019){Panamarev}, {Just}, {Spurzem}, {Berczik},
  {Wang}, \& {Arca Sedda}}]{Panamarev2019}
{Panamarev}, T., {Just}, A., {Spurzem}, R., {et~al.} 2019, \mnras, 484, 3279

\bibitem[{{Pike} {et~al.}(2021){Pike}, {Kawai}, {Mihara}, \&
  {Negoro}}]{Pike2021}
{Pike}, S., {Kawai}, N., {Mihara}, T., \& {Negoro}, H. 2021, The Astronomer's
  Telegram, 15048, 1

\bibitem[{{Sharma} {et~al.}(2019){Sharma}, {Jain}, \& {Dutta}}]{Sharma2019}
{Sharma}, R., {Jain}, C., \& {Dutta}, A. 2019, \mnras, 482, 1634

\bibitem[{{Tagawa} {et~al.}(2020){Tagawa}, {Haiman}, \& {Kocsis}}]{Tagawa2020}
{Tagawa}, H., {Haiman}, Z., \& {Kocsis}, B. 2020, \apj, 898, 25

\bibitem[{Walsh {et~al.}(2015)Walsh, Cackett, \& Bernardini}]{Walsh2015}
Walsh, A.~R., Cackett, E.~M., \& Bernardini, F. 2015, \mnras, 449, 1238

\bibitem[{{Yan} \& {Yu}(2015)}]{Yan2015}
{Yan}, Z., \& {Yu}, W. 2015, \apj, 805, 87

\bibitem[{{Ye} {et~al.}(2021){Ye}, {Kremer}, {Rodriguez}, {Rui}, {Weatherford},
  {Chatterjee}, {Fragione}, \& {Rasio}}]{Ye2021}
{Ye}, C.~S., {Kremer}, K., {Rodriguez}, C.~L., {et~al.} 2021, arXiv e-prints,
  arXiv:2110.05495

\end{thebibliography}

\end{document}